\documentclass[aps,prl,twocolumn,preprintnumbers,amsmath,amssymb]{revtex4-1}
\usepackage{graphicx}
\usepackage{epsfig}
\usepackage{dcolumn}
\usepackage{bm}
\usepackage{ulem}
\usepackage{color}
\usepackage{float}

\usepackage[colorlinks,citecolor=blue,urlcolor=blue]{hyperref}

\def\be{\begin{equation}}       \def\ee{\end{equation}}
\def\bea{\begin{eqnarray}}      \def\eea{\end{eqnarray}}
\def\ba{\begin{array}}
\def\ea{\end{array}}
\def\bnum{\begin{enumerate} }
\def\enum{\end{enumerate}}

\def\=>{\Rightarrow}
\def\>{\rightarrow}

\def\eye2{Fathbb{I}}

\def\Eq#1{Eq.~(\ref{#1})}
\def\Fig#1{Fig.~\ref{#1}}

\renewcommand{\>}{\rangle}

\begin{document}
\title{Strong and Weak Many-Body Localizations}
\author{Shi-Xin Zhang}
\affiliation{Institute for Advanced Study, Tsinghua University, Beijing 100084, China}
\author{Hong Yao}
\email{yaohong@tsinghua.edu.cn}
\affiliation{Institute for Advanced Study, Tsinghua University, Beijing 100084, China}

\begin{abstract}
\end{abstract}

\date{\today}
\maketitle

{\bf Many aspects of many-body localization (MBL) \cite{Fleishman1980, Gornyi2005, Basko2006, Oganesyan2007}, including dynamic classification of MBL phases, remain elusive.
Here, by performing real-space renormalization group (RSRG) analysis \cite{Potter2015, Vosk2015, Zhang2016, Dumitrescu2017,Zhang2018} we propose that there are two distinct types of MBL phases: ``strong MBL'' induced by quasiperiodic (QP) potential and ``weak MBL'' induced by random potential.
Strong and weak MBL phases can be distinguished by their different probability distributions of thermal inclusion and entanglement entropy: \textit{exponential} decay in strong MBL phases but \textit{power-law} decay in weak MBL.
We further discuss underlying mechanisms as well as experimental implications of having two distinct types of MBL phases.
Strong MBL induced by QP potential may provide a more robust and promising platform for quantum information storage and processing \cite{Huse2013, Choi2015, Yao2015, Smith2016}. }

Many-body localization generalizes the notion of Anderson localization \cite{Anderson1958} to interacting quantum systems \cite{Nandkishore2015, Altman2014, Vasseur2016,Abanin2017, Ponte2015,Nandkishore2017a, Yao2014, Khemani2016}.
Moreover, MBL phases violate eigenstate thermalization hypothesis (ETH) \cite{Deutsch1991, Srednicki1994, Rigol2008}, providing a unique mechanism to break ergodicity in interacting systems.
They feature various novel properties including area-law entanglement entropy for highly excited states \cite{Bauer2013}, logarithmic spread of entanglement \cite{Bardarson2012}, and emergent local integrals of motion \cite{Serbyn2013a,Huse2014}. MBL transitions separating ergodic and MBL phases \cite{Pal2010a, Vosk2013, Pekker2014,Kjall2014, Agarwal2015, Vasseur2015, Mondaini2015, Khemani2016a, Serbyn2016, Jian2017, Dai2018a, Stagraczynski2017, Thiery2017, Roeck2018} are eigenstate transitions associated with all eigenstates \cite{Luitz2015}. New paradigm beyond equilibrium quantum many-body physics has been needed to understand MBL phases and transitions. Especially, there remain many elusive aspects of MBL, including classifications of MBL phases in one dimension and the stability of MBL phases in higher dimensions.

Although it is under intense debate about the stability of MBL in two and higher dimensions, two mechanisms are known to induce MBL in 1D: one by random potential and the other by quasiperiodic (QP) potential. It has been proved in a mathematically rigorous way that MBL can exist in 1D random systems \cite{Imbrie2016}. Moreover, experimental evidences of QP-induced MBL \cite{Iyer2013, Li2015a, Modak2015, Li2016, Lee2017, Li2017a,  Chandran2017, Nag2017, BarLev2017, Gray2017, Setiawan2017, Deng2017, Crowley2018,Znidaric2018, Hsu2018, Mace2019, Xu, Doggen2019, Weiner2019} were reported recently \cite{Schreiber2015a, Bordia2016, Luschen2017, Bordia2017, Luschen2018, Lukin2019, Kohlert, Rispoli}.
One natural open question is whether randomness-induced and QP-induced MBL phases are qualitatively different or not, namely what is the general classification of MBL phases. Studying their dynamic classification can not only shed light to understanding intrinsic features of MBL but also provide important guidance of utilizing most stable MBL to protect topological edge modes at finite temperature \cite{Slagle2015, Potter2015a}.

\begin{figure}[t]
    \includegraphics[width=6cm]{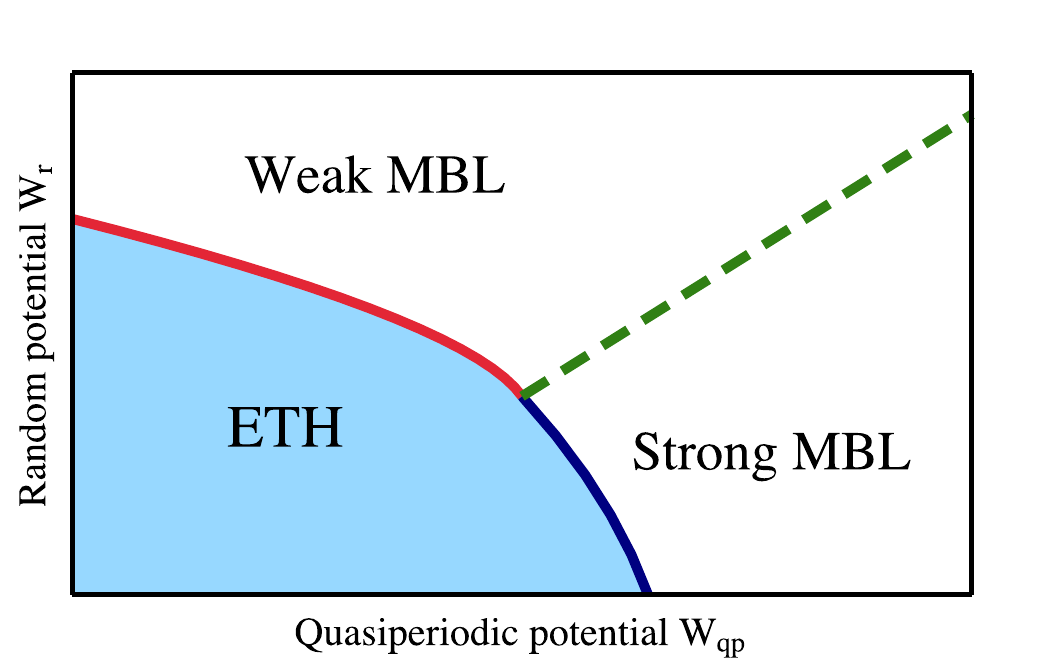}
    \caption{Schematic phase diagram of 1D interacting systems with both random potential $W_r$ and QP potential $W_{qp}$. Strong and weak MBL are two distinct types of MBL phases proposed in the present work. The transition between ETH and strong MBL features the critical exponent $\nu\approx 2.5$ while transition between ETH and weak MBL has $\nu\approx 3.2$. The dashed line represents transition between two types of MBL. }
    \label{phasediag}
\end{figure}

Here we investigate intrinsic properties of MBL phases induced by random or QP potentials from state-of-the-art RSRG with sufficiently large size $L\!\sim\!1000$.
Note that exact diagonalization (ED) is often not sufficient to extract dedicate features of MBL due to relatively small size it can access; for instance previous ED studies of MBL transitions obtained critical exponents that violate the Harris bound \cite{Khemani2016a, Khemani2017}.
We recently improved RSRG approach \cite{Zhang2018} basing on previous works \cite{Potter2015, Dumitrescu2017} by taking more microscopic details into account, which makes it especially suitable to study systems with both random and QP potentials.
By utilizing the improved RSRG, we explore systematically MBL phases induced by random and/or QP potentials and show that there are qualitative differences between randomness-induced and QP-induced MBL phases.
For randomness-induced MBL, we find that the probability distribution of half-chain entanglement entropy density features power law decay, which we call ``weak MBL''.
But, for QP-induced MBL, it follows an exponential decay, bearing the name ``strong MBL''. 
Note that the notation of strong and weak here are different from the ones in \cite{Nandkishore2014a}.
The global phase diagram in the presence of both random and QP potentials is shown in \Fig{phasediag}.

We propose that the underlying mechanism for strong (weak) MBL is the gapped (gapless) nature in the probability distribution of nearest-neighbor energy differences of single-particle states in QP (random) models.
Moreover, such intrinsic differences between two types of MBL phases can be one origin of two universality classes of MBL transitions shown in \cite{Khemani2017,Zhang2018}.
Besides, there are important experimental implications from qualitative differences between two types of MBL phases; especially it implies that the experimentally observed MBL in QP potentials can be qualitatively more robust in protecting Majorana edge modes at finite temperature.

{\bf Model and RSRG.---}We consider the following 1D interacting fermion model to investigate MBL phases:
\bea\label{model}
H=-\sum_{ij}\left(t_{ij} c^\dagger _ic_j+h.c.\right)+\sum_{i=1}^L W_i n_i+V \sum_{\langle i j\rangle}n_in_j,~~
\eea
where $c^\dag_i$ creates a fermion at site $i$, $n_i\!=\!c^\dag_i c_i$ is fermion density operator, $t_{ij}$ represents the hopping amplitude between sites $i$ and $j$, $V$ is the interaction between nearest-neighboring (NN) sites, and $W_i$ represents the onsite chemical potential which can vary from site to site.
If the hopping coefficients are restricted to uniform NN hopping, the fermion model in \Eq{model} is then equivalent to the XXZ spin model with site-varying Zeeman magnetic field. In the present work, we consider both NN hopping $t$ and next-nearest-neighbor (NNN) hopping $t'$. We consider a generic site-varying potential $W_i\!=\!W_{r,i}+W_{qp,i}$ with $W_{r,i}\!\in\![0,W_r]$ and $W_{qp,i}\!=\!W_{qp}\cos(2\pi\alpha i+\phi)$, where $W_r$ and $W_{qp}$ denote the strength of random and QP potentials, respectively, and $\alpha$ is the QP wave number (we set $\alpha\!=\!\frac{\sqrt{5}-1}{2}$). When $W_i$ has only QP part, the model represents a generalized AA model \cite{Harper1955,Aubry1980, Sarma1988,Biddle2009,Biddle2011a,Wang2013a,Ganeshan2015}. We fix $t\!=\!1$, $t'\!=\!-0.1$, and $V\!=\!0.3$ while varying $W_r$ and $W_{qp}$ in our calculations. The model in \Eq{model} can undergo a MBL transition by tuning $W_r$ or $W_{qp}$.

To study the intrinsic features of MBL phases as well as universal properties of MBL transitions in the presence of QP and/or random potentials in large scale, we utilize the improved RSRG \cite{Zhang2018} which takes more microscopic details into account and is suitable to capture the differences of MBL transitions in different microscopic settings,  e.g. random potential versus QP potential in the present work.
Usual RSRG approach \cite{Potter2015,Dumitrescu2017} mainly employ generic features of criticality, including assumptions of scaling invariance near transitions and the hierarchy of resonance clusters implemented through iterations.
The improved version takes full-featured noninteracting models into account which could provide a more suitable input basis for RSRG.
See the SM for more details.
It turns out such improvement is essential to get reliable results and provides a tool to understand the mechanism behind the two distinct universality classes.

By measuring half-chain entanglement entropy and the size of maximal thermal blocks in systems with large size, RSRG was shown to be capable of revealing critical behaviors of high accuracy in random \cite{Dumitrescu2017} and QP \cite{Zhang2018} cases.
We follow this route to study the models with both random and QP potentials.
Specifically, we investigate the probability distribution of half-chain entanglement entropy density and length of maximal thermal block across different samples (namely different disorder configurations) for both random and QP cases.
We will show in below that there exist distinct behaviors of such distributions between the two cases, which provides convincing evidences that there are two qualitatively different types of MBL phases called strong and weak MBL, respectively.

{\bf Strong and weak MBL.---}We first investigate the probability distributions of maximal thermal block $l_\textrm{max}$ normalized against the system size $L$ (namely $x\!=\! l_\textrm{max}/L$) across different samples.
For each disorder configuration (different $\phi$ for QP case), we perform RSRG iterations to get the final configuration of resonance clusters and find the size of the largest resonance cluster (or maximal thermal block). We then analyze the distribution histogram of the size of the maximum thermal block across disorder configurations. It was shown that in randomness-induced MBL phases the distribution of maximal thermal block and half-chain entanglement entropy density follows power-law statistics \cite{Dumitrescu2017, Dumitrescu2018, Goremykina2019}.
In the present work, we study the distribution behaviors for both random and QP potentials to identify possible intrinsic differences between the two cases.

\begin{figure}[t]
	\includegraphics[width=7.cm]{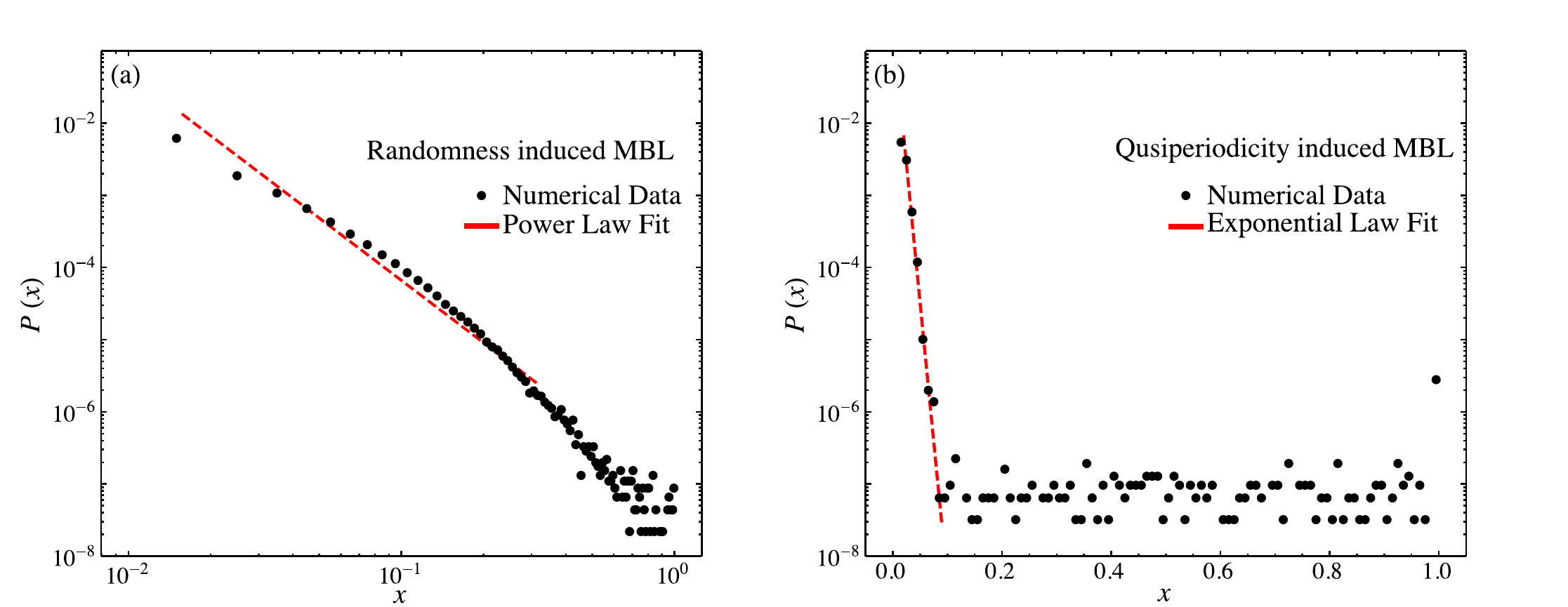}
    \caption{The probability distribution of the normalized length $x=l_\textrm{max}/L$ of maximal thermal block across samples for the random case (a) and QP case (b), respectively. The results are obtained by averaging over $10^7$ disorder configurations with $L\!=\!600$. The probability distribution follows power law in the random case (a) but exponential decay in QP case (b).}
	\label{ml2}
\end{figure}

We perform RSRG to obtain distributions of normalized maximal thermal block length $x$ across samples for random and QP cases separately, as shown in Fig. \ref{ml2}. First of all, in both cases the distribution has a flat tail with relatively low probability followed by a thermal peak with relatively high probability. 
The flat tail with relatively small probability is due to background fluctuation with rare occurrence as the number of sampling is finite. The thermal peak can also be understood as finite size effect since all thermal blocks longer than the system size theoretically contribute to the thermal peak in the histogram.
Note that the probabilities of the thermal peak and of the flat tail in the distribution decreases with increasing potential $W_r$ or $W_{qp}$ and both are smeared out deep in the MBL phases.

Therefore, to obtain essential characterization of the distribution patterns, we need focus on the intrinsic parts in the probability distributions (namely not including the flat tails and thermal peaks).
For the random potential case, this distribution follows the power law $P(x)\!\propto\!x^{-\alpha}$, where $\alpha\!\approx\! 2$ at (de)localization transition and becomes larger in the MBL phase with increasing $W_r$, as shown in Fig. \ref{ml2}(a). 
While it is numerically challenging to completely rule out stretched exponential \cite{Morningstar2019}, this power-law fitting in the weak MBL phase is consistent with previous works \cite{Dumitrescu2018,Dumitrescu2017} implying that the power-law behaviours in the weak MBL phase are quite robust.
However, the distribution of the QP case is qualitatively different from the random case and clearly shows exponential decay, i.e. $P(x)\!\propto\!e^{-\gamma x}$ where $\gamma$ is a constant depending on $W_{qp}$, as shown in \Fig{ml2}(b).
To the best of our knowledge, it is the first time that a good exponential decay in the probability distribution is clearly observed in MBL phases. We label such phases with exponential decay feature as strong MBL, which are qualitatively different from weak MBL with power-law decay.

The result of exponential distribution for the QP-induced MBL phase obtained from RSRG is of high importance in several aspects.
Firstly, due to the limitation of the method itself, it was previously thought that RSRG approach cannot give rise to exponential distribution. The result presented here for the QP case shows the effectiveness of RSRG in a broader region.
Secondly, the distinct distribution between the QP-induced and randomness-induced MBL phases is a strong sign that the two MBL phases are intrinsically different.
After accepting that there are two distinct types of MBL phases, it would be straightforward to understand the two universality classes of MBL transitions. Namely, the distinction in the universality classes of MBL transitions between QP and random cases is due to the two MBL phases themselves are qualitatively different.

More importantly, the different distributions may further imply differences in the `robustness' of MBL phases.
As the distribution of maximal thermal block across different samples is qualitatively the same as the distribution of all underlying thermal blocks of a single sample in the thermodynamic limit, the exponential decay shown in the QP-induced MBL phases would indicate that the rare thermal region in QP-induced MBL phase is formed by independent events.
Namely, the probability of forming a thermal block with size $x$ in QP-induced MBL phases should be given by $P(x)\propto p^{-x}\propto e^{-\gamma x}$, where $p$ is the probability of thermalization on one site.
For the power-law distribution observed in random-induced MBL phases, it implies that the thermalization on each site is not independent but is strongly coordinated in some sense.
That is to say, a system with power law distribution of entanglement entropy or maximum thermal block should be easier to thermalize when coupled to an external thermal bath, since the thermalization of one site increases the probability of thermalization on other sites.
In other words, the hidden correlation between rare-region thermalization events renders random-induced MBL phases more fragile against the thermal bath than QP-induced MBL phases where thermalization is an independent event for each site.
This is why we label the QP-induced (random-induced) MBL phases as strong (weak) MBL.
Such strong MBL could be more suitable for quantum information processing \cite{Huse2013, Choi2015, Yao2015, Smith2016}.

\begin{figure}[t]
	\includegraphics[width=3.5cm]{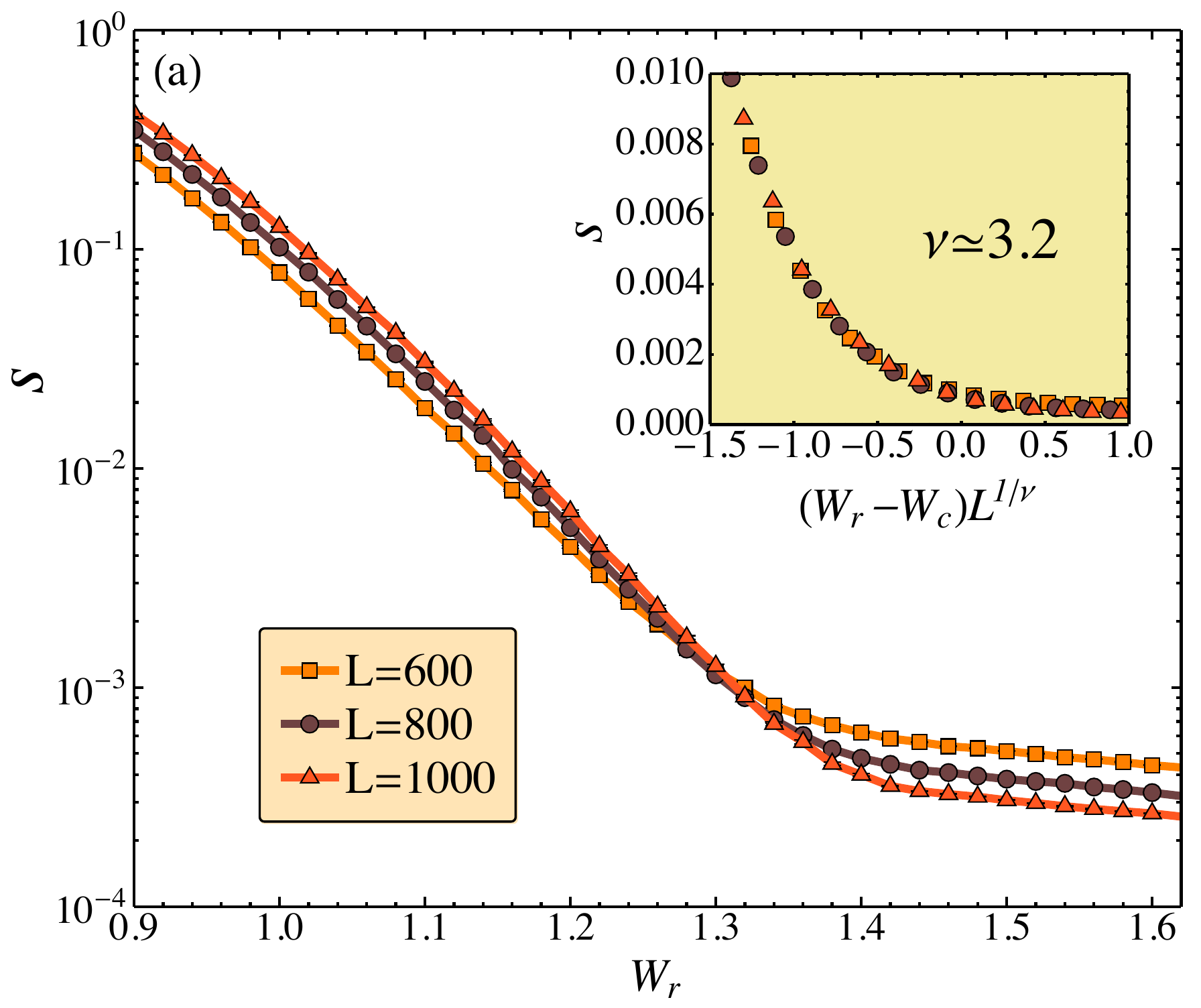}~~~
    \includegraphics[width=3.5cm]{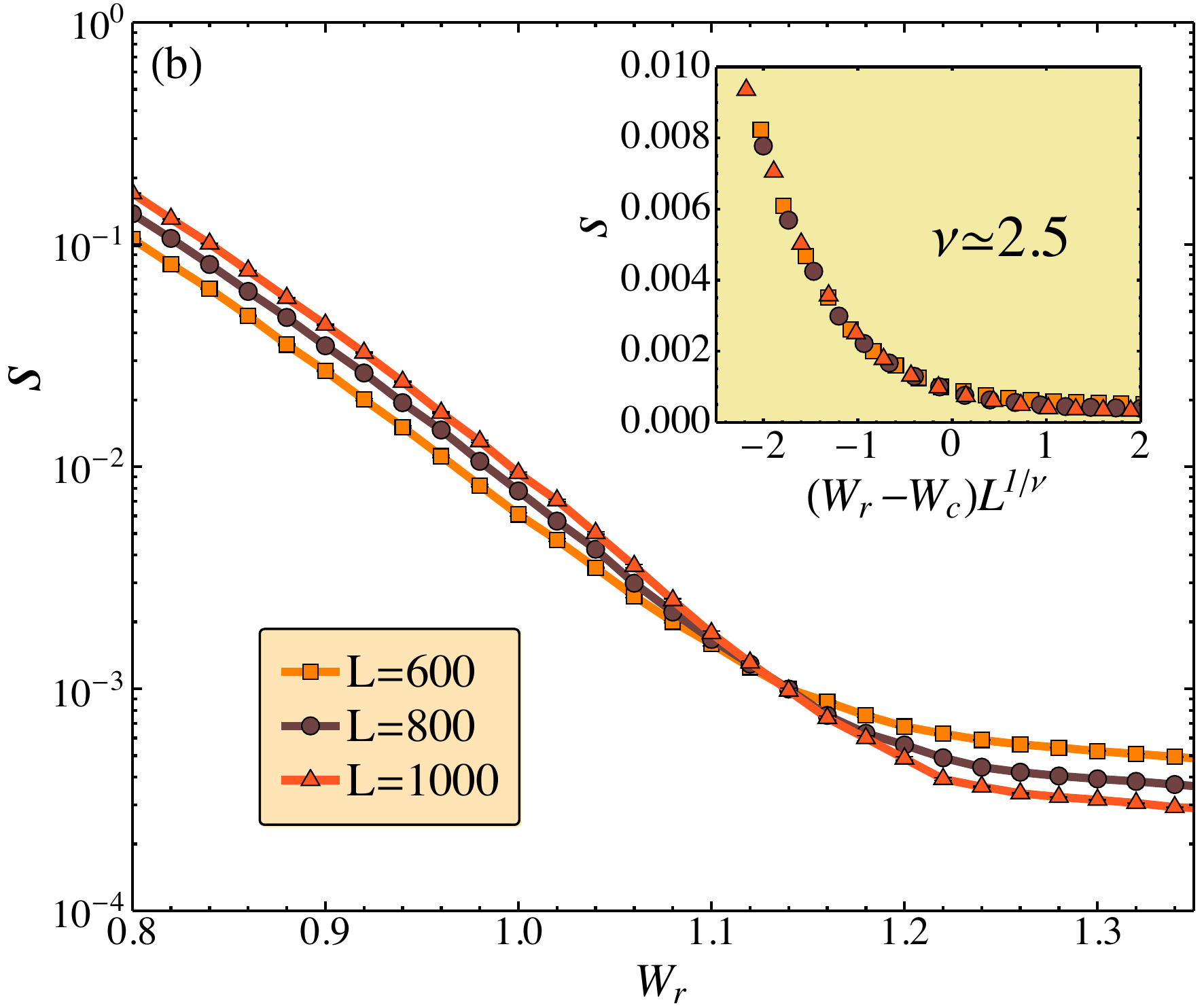}
    \caption{The half-chain entanglement entropy density as a function of random potential $W_r$ and fixed QP potential $W_{qp}$. The data collapse is shown in the inset. (a) For $W_{qp}\!=\!2.3$, we obtain the critical random potential $W^c_r\!=\!1.33\!\pm\! 0.01$ and the critical exponent $\nu\!=\!3.2\!\pm\! 0.3$. (b) For $W_{qp}\!=\!2.5$, we obtain $W_c^r\!=\!1.15\!\pm\! 0.01$ and $\nu=2.5\!\pm\! 0.3$. }
	\label{qp2}
\end{figure}

{\bf The global phase diagram.---}We further perform RSRG calculation on models with mixed random and QP potentials to provide more evidences of two distinct types MBL phases.
By calculating the half-chain entanglement entropy density across different disorder configurations with various system size and varying random potential $W_r$ for a given finite QP potential $W_{qp}$, we can locate MBL transitions in the presence of both types of potentials by the crossing of the half-chain entanglement entropy density with different system size $L$.
From the scaling $s\equiv S/L=f((W_r-W^c_r)L^{1/\nu})$ near the critical point, we can extract the critical exponent $\nu$ by finite-size scaling analysis.
We obtain $\nu\!\approx\!3.2$ for $W_{qp}\!<\!2.4$ while $\nu\!\approx\! 2.5$ for $W_{qp}\!>\!2.4$. Typical results are shown in Fig. \ref{qp2}.

Note that the critical exponents we extracted here in the presence of mixed potentials are consistent with the result of either pure random or pure QP potential \cite{Zhang2018}. These results support the proposal of two distinct types of MBL phases and two distinct universality classes of MBL transitions. The global phase diagram with both types of potentials are sketched as \Fig{phasediag}. There is a `multicritical point' where ETH, strong MBL, and weak MBL meet. The phase diagram also supports that the purely QP-induced MBL transition is robust against small randomness and hence Harris stable with $\nu\!>\!2$.

We further investigate the distribution of physical quantities such as half-chain entanglement entropy across samples in the presence of mixed potentials. For $W_{qp}\!=\!2.2$ and $W_r=1.48$, it is clearly in power law, which is the same to the pure random limit. On the contrary, for $W_{qp}\!=\!3.1$ and $W_r=0.74$, the result is similar to the pure QP case.
This is another important evidence supporting the proposal for two distinct types of MBL phases.
Note that there should be a phase transition line (denoted as dashed line in \Fig{phasediag}) between the strong MBL and weak MBL phases.
Nonetheless, it is numerically challenging to determine the precise locations of transitions between the two types of MBL phases as there is a narrow region in the parameter space where the distribution is not clearly in power law or in exponential decay. The dash line in the phase diagram is an illustration.

\begin{figure}[t]
	\includegraphics[width=7cm]{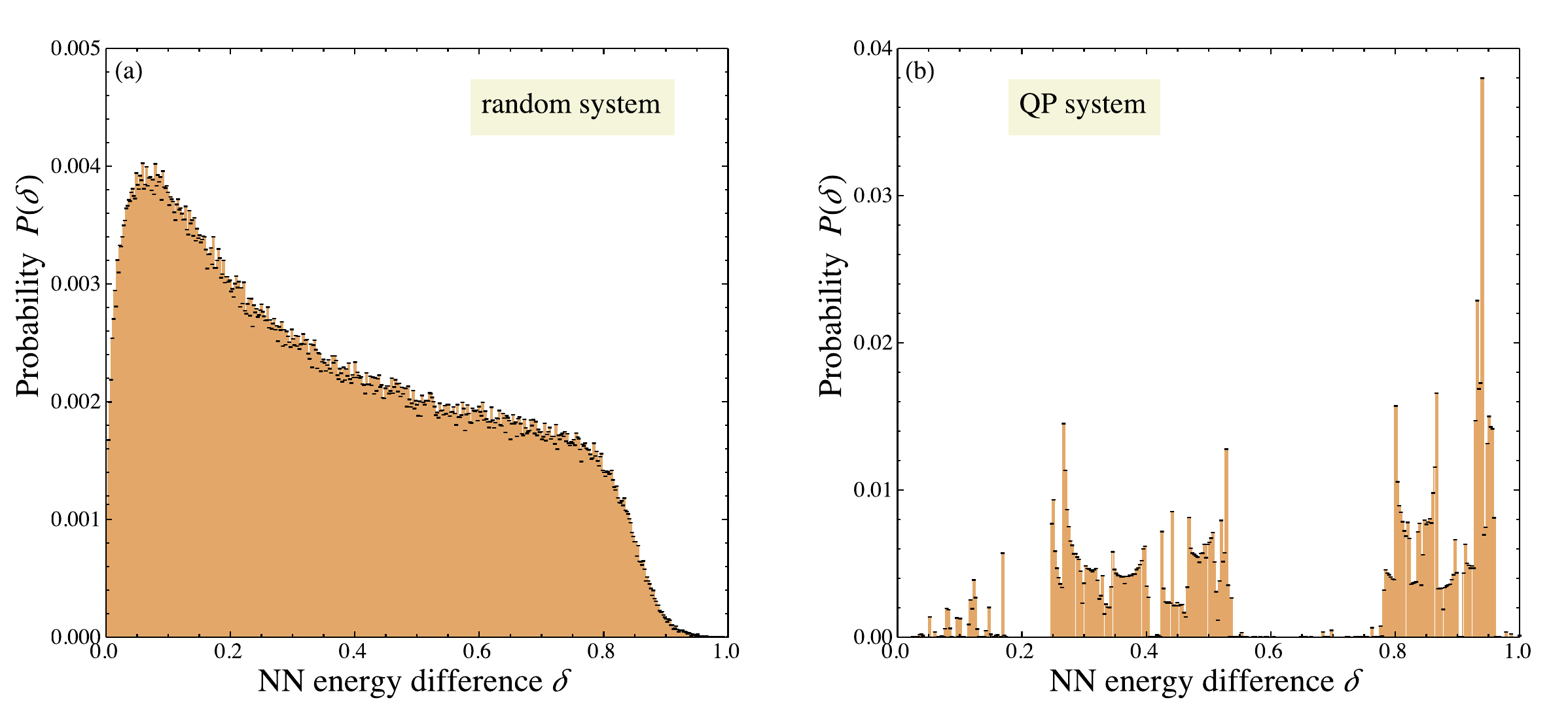}
    \caption{Probability distribution of NNED obtained by averaging over 1000 disorder configurations near MBL criticality with $L\!=\!600$ in weak MBL phase (a) and strong MBL phase (b), respectively. (a) The NNED probability is gapless for random case. (b) The NNED probability near $\delta\!=\!0$ is vanishing, indicating NNED probability gap for QP case.}
	\label{gaphist}
\end{figure}

{\bf Discussions and conclusions.---}We have shown convincing evidences that there are two distinct types of MBL phases, which underlies two different universality classes of MBL transitions.
Here we provide an explanation on why there are two types of MBL phases, one of which is qualitatively stronger than the other.
The qualitative difference between two types of MBL phases may originate from their different features in the nearest-neighbor energy difference (NNED) in single-particle states as we explain below.
It is generally believed that MBL phases emerge from Anderson localization of the noninteracting limit upon turning on interactions.
In the non-interacting limit, each single-particle state is characterized by its energy $\epsilon_n$ and localization center $r_n$, where $n$ is ordered according to the localization center position ($r_1\!<\!r_2\!<\!\cdots\!<\!r_L$).
Both spectra and localization centers of single-particle states are important information since they are set as initial input data of the RSRG iterations.
By computing NNED $\delta_{n}\!=\!|\epsilon_n-\epsilon_{n+1}|$ for different disorder configurations of a given microscopic model, one can then plot the probability distribution $P(\delta)$.
For the case of random potential, the NNED is gapless, as shown in \Fig{gaphist}(a). However, for the case of QP potential, NNED shows a  gap in the probability density $P(\delta)$, which is called NNED gap, as shown \Fig{gaphist}(b).

The above intrinsic difference in NNED between random and QP models implies that there are far more sites with close energy to their neighbors in the random case than in the QP case.
In other words, the probability that neighboring localized single-particle states have similar energy is surprisingly low in the QP case.
Since in RSRG analysis neighboring sites with closer energy is easier to form thermal clusters, probability of getting larger thermal blocks is higher if the probability distribution of NNED is gapless near $\delta\!=\!0$.
Gapless probability of NNED could cause some hidden correlation of thermalization event on each site, rendering power law distribution of maximal thermal blocks.
For the QP potential, since NNED probability distribution has a gap near $\delta\!=\!0$, the probability of forming resonance clusters at the beginning of RSRG iterations is already exponentially suppressed due to the gap. Consequently, it is more robust against thermalization and renders an exponential decay in the distribution of maximal thermal blocks.

To verify the explanation above, we design a special 1D fermion two-band model with random potential such that its single-particle eigenstates in some parameter region may have NNED gap which is similar with gap feature of QP models. (See the SM for more details on the special two-band model.) We then perform RSRG calculations on the two-band model for different parameters with and without a NNED probability gap, respectively. Indeed, we successfully observed the two distinct universality classes of MBL transitions in this two-band random model, indicating that NNED probability gap should play a key role in causing two different types of MBL phases, one of which is more robust than the other.

The identification of strong and weak MBL phases is of high experimental relevance.
The QP potential has been realized in optical lattices experimentally, in which MBL phases of atoms have been observed \cite{Schreiber2015a}. Such MBL phases induced by QP potential are strong MBL.
Random potential can be realized by speckle potentials or by programmable quantum qubits \cite{Smith2016}.
Measurements of critical exponents in the two cases could lay a solid test on the distinctions between two types of MBL phases or MBL transitions.
Furthermore, by directly coupling the prepared MBL systems with an external thermal bath
and then measuring the robustness of MBL phases against thermalisation (such as measuring the relaxation dynamics of charge imbalance), qualitative differences in the strong and weak MBL phases could be revealed.

In conclusion, we have shown convincing evidences of two distinct types of MBL phases, which are characterized by qualitatively different distributions of quantities such as entanglement entropy and possess distinct robustness against thermalization. The existence of two types of MBL phases underlies the scenario of two universality classes of MBL transitions. The global phase diagram in the presence of both types of potential is obtained. Moreover, we propose possible mechanism behind the differences between two distinct types of MBL phases and discussed their experimental implications. We believe that our findings could not only shed new light to understanding intrinsic nature of thermalization and many-body localization but also provide guides of designing more robust MBL phases that provide promising platform for quantum information storage and processing.

{\it Acknowledgement.---}We would like to thank Sarang Gopalakrishnan, David Huse, Vedika Khemani, and Romain Vasseur for helpful discussions.
This work is supported in part by the NSFC under Grant No. 11825404 (S.-X.Z. and H.Y.), the MOSTC under Grant Nos. 2016YFA0301001 and 2018YFA0305604 (HY), and the Strategic Priority Research Program of Chinese Academy of Sciences under Grant No. XDB28000000 (HY).

\begin{widetext}
\section{Supplemental Materials}
\renewcommand{\theequation}{S\arabic{equation}}
\setcounter{equation}{0}
\renewcommand{\thefigure}{S\arabic{figure}}
\setcounter{figure}{0}
\subsection{A. Details of the model}
Now we discuss further details on the 1D spinless fermion model in the main text which is utilized as the platform of RSRG calculation. In the model, $W_{r,i}$ is the random potential with uniform distribution $W_{r,i}\!\in\![0,W_r]$ for the random case while a cosine potential $W_{qp,i}\!=\!W_{qp}\cos(2\pi \alpha i+\phi)$ with incommensurate wave vector $\alpha$ and phase $\phi$ is the QP part. For all the calculations discussed in the main text of the present work, we set the irrational number as the golden ratio: $\alpha\!=\!\frac{\sqrt{5}-1}{2}$. Actually, we have also carried out similar real-space RG calculations on QP models with other wave vectors $\alpha$, such as $\alpha=\sqrt{2}/2$ and $(\sqrt{3}-1)/2$, and obtained results of critical exponents which are are consistent with the case of $\alpha=\frac{\sqrt{5}-1}{2}$ discussed in the main text.

In the limit of vanishing interaction $V\!=\!0$, the system with random potential is in the Anderson localization (AL) phase for any finite randomness $W_r\!>\!0$.
While for the QP case, the noninteracting version of the model with only NN hopping is so-called Aubry-Andr\'e (AA) model.
The AA model provides an example of single-particle localizations in 1D with finite critical potential strength but without single-particle mobility edge (SPME). There have been various generalizations of the AA model which in general possess SPME. This fact indicates that the original AA model with only NN hopping is special and not generic in the family of QP models.
To mimic the experimental settings and realistic scenarios, both NN hopping $t$ and NNN hopping $t'$ need be considered such that the model is capable of describing more generic systems.

For both type of potentials, if the noninteracting model is deep in the localized phase, it is expected that such localization can persist in a range of finite interaction $V>0$, which is known as MBL.
The transition and universal critical behavior between MBL and thermal phases are very important to understand the process of thermalization and the formation of MBL. There are reported evidences that the critical exponents of such many-body (de)localization phase transitions are different between random case and QP case, indicating two distinct university classes of MBL transitions. Trying to understand the underlying mechanism for the differences is the focus of the present study.

\subsection{B. Technical details on RSRG}
In general, RSRG develops the thermal block structures in a give system step by step according to some iterative process proposed based on physics rules.
We compare the energy difference and the tunneling strength between each blocks and determine how to merge them into larger thermal clusters in each step.
The fix point of RSRG for each disorder configuration is a system with thermal block distribution information.
For example, we know about how many thermal blocks and how large are they in the given model Hamiltonian and disorder configurations.
Note the size of the maximal thermal block is a very important physical observable directly related to the many-body (de)-localization transitions.
Besides, we can also extract the half-chain entanglement entropy information from the thermal cluster structure.
The expectation value and fluctuation distribution of the above two observables provide us valuable insights on MBL phases and transitions.

\subsection{C. Distribution data on normalized entanglement entropy}
The probability (or histogram) distribution of half-chain entanglement entropy density in both randomness-induced and QP-induced MBL phases are shown as \Fig{eedist}. It is clear that qualitative features of the entanglement entropy distribution are consistent with the probability distribution of the normalized maximal thermal block, as mentioned in the main text.

\begin{figure}
    \includegraphics[width=0.5\textwidth]{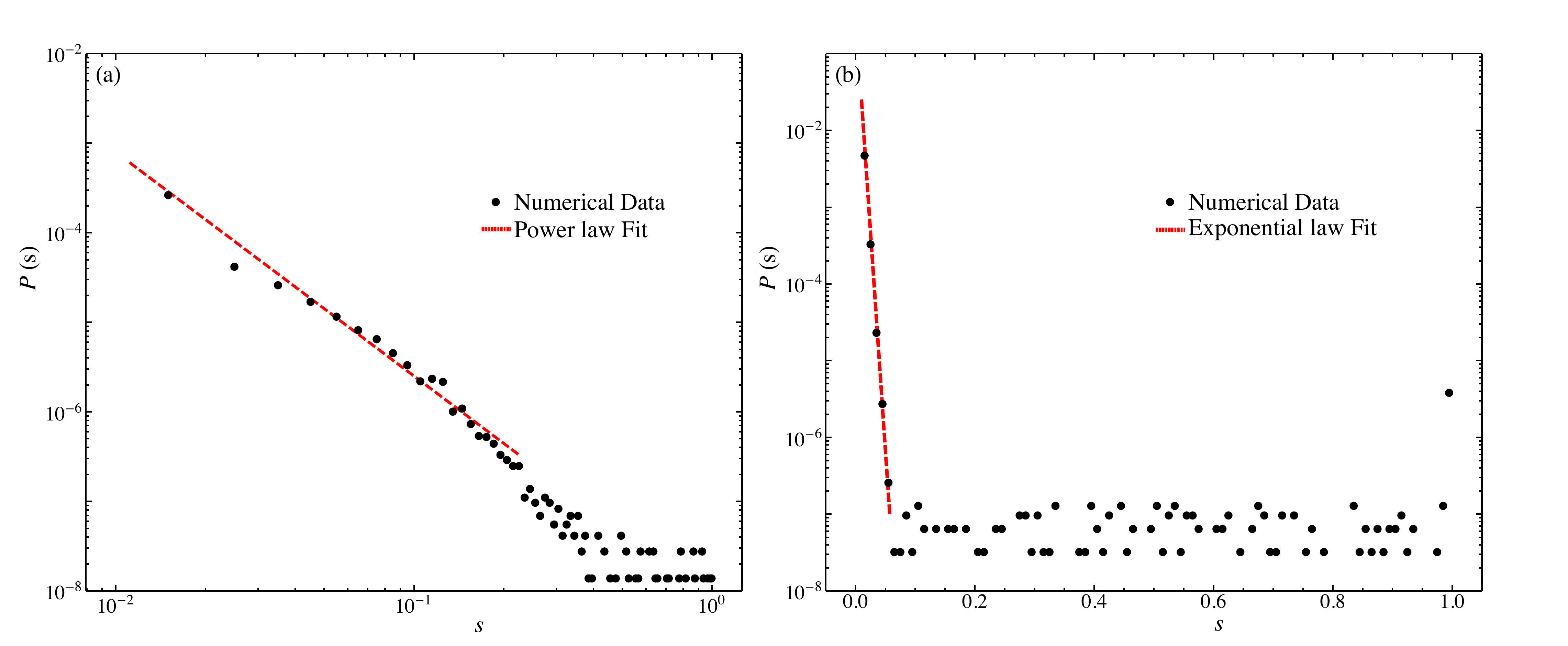}
    \caption{Probability (or histogram) distribution of normalized half-chain entanglement entropy in random-induced MBL system (a) and QP-induced MBL system (b). (a) The distribution across different disorder configurations follows power law. (b) The probability distribution decays exponentially. The distribution of entanglement entropy density are qualitatively consistent with the distribution of maximal thermal block shown in the main text.}
    \label{eedist}
\end{figure}

Here we would like to make some comments on the fitting results of distribution histogram. We think that it is quite apparent that the probability distribution of entanglement entropy or thermal inclusion in the randomness-induced MBL shows power-law behavior. Firstly, from the consistent fitting method and selection of fitting range, we obtained the power-law exponent $\alpha\approx 2$ at the MBL transition point, which is consistent with the theoretical prediction \cite{Dumitrescu2018}. Note that the weak MBL phase in our work shares similar features with the MBL* phase investigated in \cite{Dumitrescu2018}. Moreover, a number of previous studies with different RG schemes give rise to power-law distributions which are consistent with our results and shows the robustness of power-law results. We also note that the quality of power-law fitting is somewhat affected (i.e. curved down) for intermediate size of thermal inclusion due to the crossover from pure power-law to noisy flat tails. Finally, while it is numerically challenging to completely rule out stretched exponential decay suggested in a recent study with different RG scheme \cite{Morningstar2019}, our results are far better fitted with a power law. Moreover, a stretched exponential is still quite distinct from the exponential decay observed in the QP case. In other words, our main conclusion about the two distinct types of MBL phases with qualitatively different distribution holds no matter whether the distribution in the randomness-induced MBL features power-law or stretched exponential.

\begin{figure}[t]
	\includegraphics[width=0.37\linewidth]{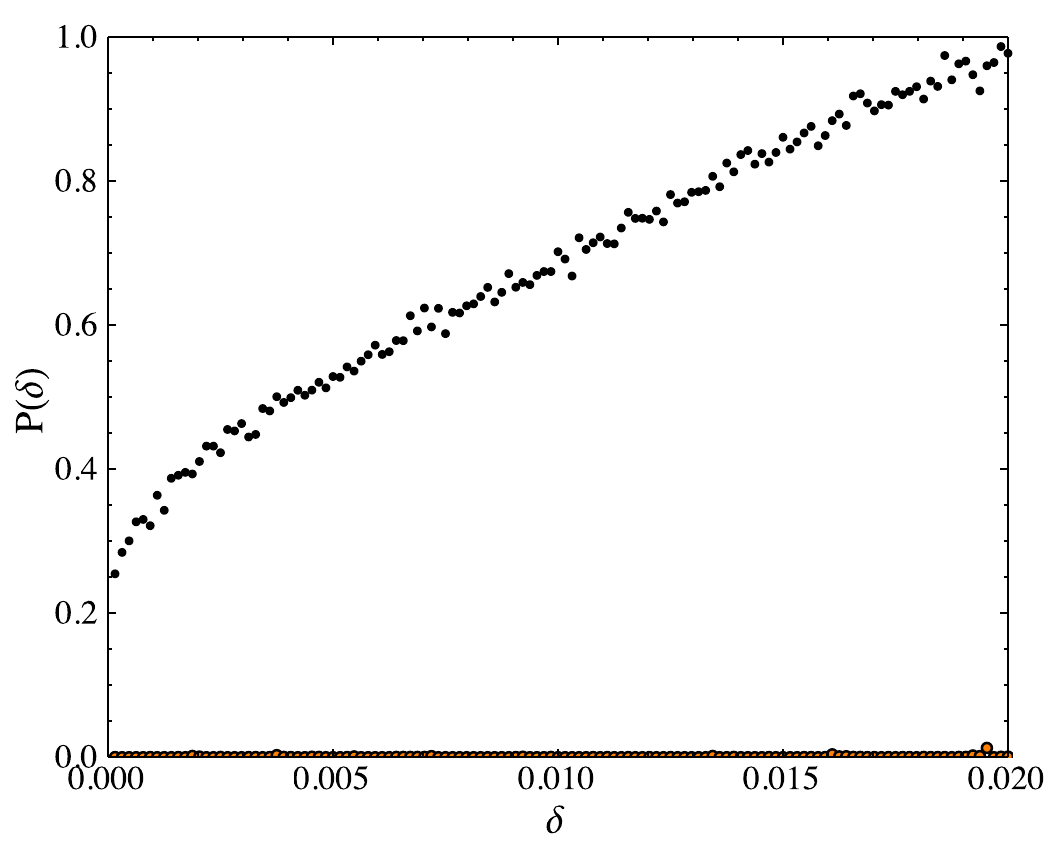}
    \caption{The zoom-in comparison of NNED probability distribution between the random and QP cases. The energy axis ($\delta$-axis) has been normalized such that the maximal energy difference is set to 1. The data is collected from random or QP 1D model with size $L=6400$. The calculation is carried out with $1800$ phase configurations in QP model and $4500$ disorder configurations in random model. There are $6400$ bins in total, and only the first 128 bins (i.e. $\delta\leq 2\% \;\delta_{max}$) are shown in the figure.
    The black dot is for the case of random potential while the orange dot is for the QP case. It is clear the NNED probability distribution of the random-potential model remains finite around $\delta=0$ while it clearly shows probability gap with rare noise in the model with QP potential.}
	\label{nnedcomp}
\end{figure}

\subsection{D. NNED spectrum near zero energy}
As mentioned in the main text, NNED probability distribution for the case of random potential is gapless near $\delta=0$. However, due to the hybridization of neighboring eigenstates, there are less states with small energy difference which can be observed in the drop of probability in the leftmost part of Fig. 4(a) in the main text. Namely, there is energy level repulsion in the system. Nonetheless, the NNED probability distribution is gapless even with such level repulsion. To show this, we zoom in the spectrum around zero energy difference and compare the random case with the QP one in the same figure, as shown in Fig. \ref{nnedcomp}. The gapless NNED spectrum in random case is apparently distinct from the QP case where it features vanishing probability close to zero energy difference.

We think that the difference in the NNED spectrum is the key reason for two distinct types of MBL phases, which is further supported by the results on two-band random model as given by Sec. E below. Moreover, the existence of such gap for the QP case may also be the reason underlying the Harris robustness of QP-induced MBL transition. Since small on-site randomness cannot destroy the NNED gap in the QP case, the universality class for QP-induced MBL transition with weak quenched randomness is expected to remain the same as the pure QP case.

\subsection{E. Two universality classes results on two-band random model}
We now discuss the details of the special two-band model with random potential. It is 1D interacting fermion chain with a specially chosen \textit{random} potential $W_i$. Specifically, $W_i=W_{1,i}$ with $W_{1,i}\in [0,W_1]$ uniformly or $W_i=W_{2,i}$ with $W_{2,i}\in [W_1+\Delta,W_1+\Delta+W_2]$ uniformly. Namely, the potential $W_i$ lies in either the potential segment $[0,W_1]$ or the segment $[W_1+\Delta,W_1+\Delta+W_2]$. The two potential ranges with width $W_1$ and $W_2$, respectively, are separated by a gap $\Delta$.
Suppose that the probability for two NN sites having $W_i$ in the same segment is $p$. Namely, the probability of potential $W_{i+1}$ in the same potential segment with $W_i$ is $p$.
For simplicity, we consider the limit of vanishing $t$ and $t'$.
If $p$ is sufficiently small, the special random model has a `soft' NNED probability gap, similar as the QP model, because the neighboring sites have a large probability that their potentials sit on different potential segments.
However, when $p=0.5$ the NNED probability distribution is gapless such that this two-band random model has no quantitative difference from the usual random model.

We performed RSRG calculations on such a two-band random model with $p=0.5$ and $p=0.003$, respectively.
The half-chain entanglement entropy density is shown as Fig. \ref{tbands} for various system size $L$.
The critical exponent for the case of $p=0.003$ is indeed around $\nu\approx 2.5$, which is the same as the QP case. The critical exponent $\nu\approx 3.2$ for the case of $p=0.5$, similar with the result of the usual random model.
The results obtained for this two-band random model give decesive support to our proposal that the NNED probability gap is the mechanism behind two different universality classes of MBL transitions and two distinct types of MBL phases. These results above imply that strong MBL phases can also be achieved by specially designed random potentials although it might be experimentally challenging to design such random potentials whose spectra have the so-called NNED probability gap. In other words, it is more straightforward to employ QP potential to achieve strong MBL phases.

\begin{figure}[ht]
    \includegraphics[width=0.6\linewidth]{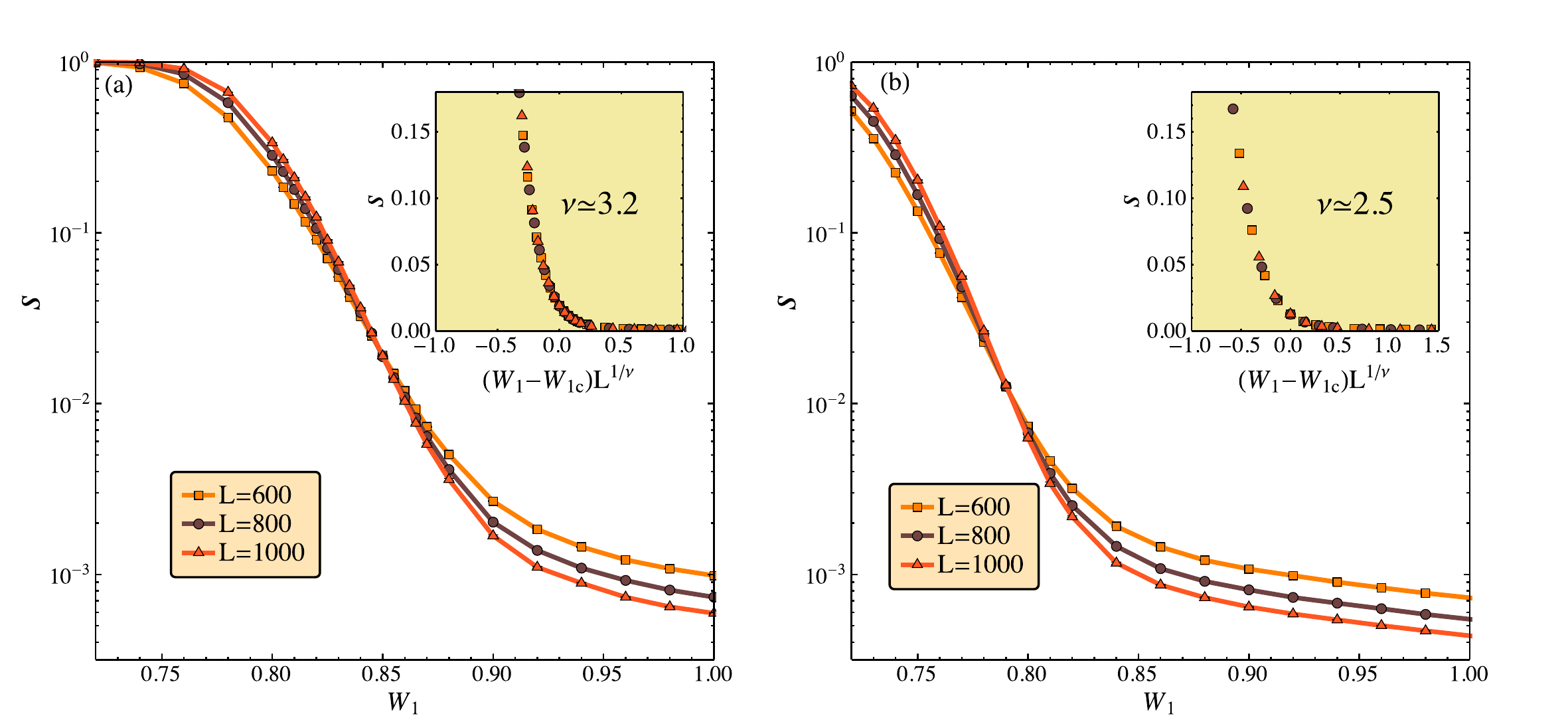}
    \caption{The half-chain entanglement entropy density of different system size near MBL criticality in the specially designed two-band random model. The random potentials of the model have two bands with band width $W_1$ and $W_2$, respectively, which are separated by a gap $\Delta$. We set $W_1=W_2=2\Delta$. The probability of having potentials of two NN sites in the same band is $p$. (a) For $p=0.5$, the critical point is given by $W_{1c}=0.85\pm0.01$, and the critical exponent of such MBL criticality is estimated as $\nu=3.2\pm 0.3$. (b) For $p=0.003$, the critical point is given by $W_{1c}=0.79\pm0.01$, and the critical exponent of such MBL criticality is estimated as $\nu=2.5\pm 0.3$. This critical behavior is consistent with the case of QP potential.}
	\label{tbands}	
\end{figure}

\end{widetext}

\end{document}